\documentclass[aps,prl,twocolumn,superscriptaddress]{revtex4-1}

\usepackage{graphicx}
\usepackage{amssymb}
\setcitestyle{super}

\begin{document}


\title{Critical Velocity in the Presence of Surface Bound States in Superfluid $^3$He-B}


\author{P. Zheng}
\affiliation{Department of Physics, University of Florida, Gainesville, Florida 32611-8440, USA}
\author{W.G. Jiang}
\affiliation{Department of Physics, University of Florida, Gainesville, Florida 32611-8440, USA}
\author{C.S. Barquist}
\affiliation{Department of Physics, University of Florida, Gainesville, Florida 32611-8440, USA}
\author{Y. Lee}
\email[]{yoonslee@phys.ufl.edu}
\affiliation{Department of Physics, University of Florida, Gainesville, Florida 32611-8440, USA}
\author{H.B. Chan}
\affiliation{Department of Physics, The Hong Kong University of Science and Technology, Clear Water Bay, Kowloon, Hong Kong}


\date{\today}

\begin{abstract}
A microelectromechanical oscillator with a gap of 1.25~$\mu$m was immersed in superfluid $^3$He-B and cooled below 250~$\mu$K at various pressures. Mechanical resonances of its shear motion were measured at various levels of driving force. The oscillator enters into a nonlinear regime above a certain threshold velocity. The damping increases rapidly in the nonlinear region and eventually prevents the velocity of the oscillator from increasing beyond the critical velocity which is much lower than the Landau critical velocity. We propose that this peculiar nonlinear behavior stems from the escape of quasiparticles from the surface bound states into the bulk fluid.
\end{abstract}


\maketitle

Superfluidity is associated phenomenologically with dissipationless flow of fluid, although its physical implications are more profound. It is a common phenomenon occurring in all classes of quantum gases and fluids, whether they are charged or neutral, bosonic or fermionic. Consider a uniform superfluid at zero temperature flowing through a narrow channel. As the velocity increases gradually, the flow becomes dissipative above a threshold velocity due to the energy loss in generating excitations in the layers of fluid close to the walls \cite{Landau1941JP1}. Landau first recognized this mechanism and derived the so-called Landau critical velocity, $v_{_L} = min\{E(p)/p\}$. Here, $E(p)$ represents the dispersion of the excitation. Those excitations are phonons in Bose-Einstein Condensation of cold atoms, rotons in superfluid $^4$He (He\,II), or Bogoliubov quasiparticles in fermionic superfluid $^3$He and superconductors. The Landau criteria have been experimentally verified in many systems using moving objects in a static host \cite{Ahonen1976PRL1, Allum1977PTRSA1, Weimer2015PRL1, Raman1999PRL1}. Specifically, in He\,II, $v_{_L}=\sqrt{2\Delta_r/\mu}\approx 60$~m/s with the roton spectrum, $E(p)=\Delta_r+p^2/2\mu$, while $v_{_L}\approx \Delta/p_{_F}\approx 60$~mm/s in the B-phase of superfluid $^3$He at around 20 bar with an isotropic gap, $\Delta$, and Fermi momentum, $p_{_F}$. However, this simple picture is often complicated in reality by various mechanisms such as vortex pinning in superconductors \cite{DewHughes2001LTP1} and nucleation of quantized vortices in superfluids, producing a wide range of critical velocities particularly in He\,II \cite{Varoquaux2006CRP1, Raman1999PRL1}.

Superfluid $^3$He, a prime example of unconventional Cooper pairing, may present a rather unique complication in this respect. Pair-breaking by scattering from any type of disorder or impurity is an exceptional feature of unconventional pairing with a non-zero angular momentum \cite{Abrikosov1961JETP1, Larkin1965JETPL1}. Undoubtedly, interfaces and surfaces also serve as effective pair-breaking agents, resulting in sub-gap bound states spatially localized near the surface within the coherence length, $\xi_0$, called the surface Andreev bound states (SABS) \cite{Buchholtz1981PRB1, Ambegaokar1974PRA1, Zhang1987PRB1, Nagato1998JLTP1, Vorontsov2003PRB1}. The B-phase of superfluid $^3$He has an isotropic gap, a rare case for p-wave pairing. It is also known to be a 3D time-reversal invariant topological superfluid \cite{Mizushima2015JPCM1, Schnyder2008PRB1}. Therefore, the SABS of $^3$He-B are topological excitations emerging from the bulk-edge correspondence that are theoretically predicted to host Majorana fermions \cite{Chung2009PRL1}. An interesting question naturally arises: What is the role of the SABS in the dissipation mechanism of flow near a boundary? In a recent experiment by the Lancaster group \cite{Bradley2016NP1}, the researchers found that their wire moving with a constant speed behaved unexpectedly: no critical velocity was observed, even at velocities exceeding $v_{_L}$. They argued that the presence of surface states would isolate the bulk from the motion of an object and consequently shuts down the Landau process mentioned above.

In this Letter, we report an unusually low critical velocity in $^3$He-B above which a massive amount of quasiparticles are generated. We believe that this behavior is directly related to the microscopic structure of SABS near a diffusive boundary, and consistent with our recent report on the anomalous low temperature dependence of the damping of the MEMS oscillator \cite{Zheng2016PRL1}.

In this work we employed a mechanical oscillator which was developed specifically to investigate the phenomena related to the surface states in a confined geometry \cite{Gonzalez2013RSI1}. Different types of mechanical oscillators such as vibrating wires \cite{Guenault1986JLTP1, Yano2005JLTP1}, tuning forks \cite{Blaauwgeers2007JLTP1}, vibrating grids \cite{Bradley2008PRL1, Bradley2012PRB2}, and nanoeletromechanical wires \cite{Defoort2016JLTP1} have been successfully exploited in both superfluid $^3$He and $^4$He. However, our oscillator possesses several features that are advantageous for this purpose. Our microelectromechanical system (MEMS) based oscillator is composed of a moving plate with a high aspect ratio. This geometry maximizes the coupling between the oscillator and the surface states. The oscillating plate moves in the direction of the plane in the shear mode as a whole without a velocity gradient. These devices have been successfully used in studying normal \cite{Gonzalez2016PRB1} and superfluid $^3$He \cite{Zheng2016PRL1} in the linear regime where the oscillator velocity is relatively low.

The MEMS device used in this measurement has a 2~$\mu$m thickness mobile plate with 200~$\mu$m lateral size. The plate is suspended above the substrate by four serpentine springs, maintaining a gap of 1.25~$\mu$m. When the device is submerged in a fluid, a film is formed inside the gap, while the bulk fluid is in direct contact with the top surface of the plate. The details of the measurement technique can be found elsewhere \cite{Gonzalez2013RSI1, Barquist2014JPCS1, Zheng2016JLTP1}.
\begin{figure}
\includegraphics[width=0.8\linewidth]{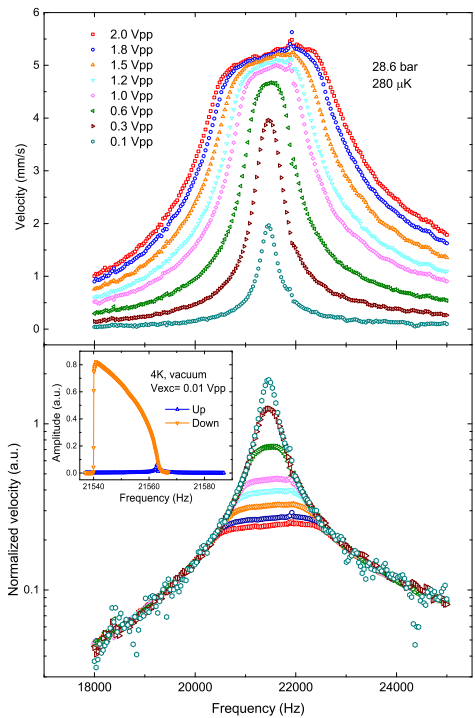}
\caption{Resonance spectra of the oscillator in superfluid at various excitations at 28.6~bar and 280~$\mu$K in 14~mT. (\textit{Top}) Resonance spectra without normalization. (\textit{Bottom}) Normalized resonance spectra by the excitation in a semi-log scale. The glitches at 22~kHz are instrumental artifacts. (\textit{Inset}) Nonlinear resonance spectrum at 4~K in vacuum. \label{NLspectra}}
\end{figure}

In this experiment two methods were adopted for the measurements of the MEMS in fluid. One is the frequency sweep where a spectrum is obtained by sweeping the driving frequency through the resonance ($\approx 20$~kHz) of the shear mode with a fixed excitation. The other is the excitation sweep where the driving force is stepped upwards and downwards while the frequency is kept at the resonance where the driving force balances out the damping force. Using this method a velocity-force relation can be acquired. 

The MEMS device was cooled in liquid $^3$He to a base temperature of about 250~$\mu$K at pressures of 9.2, 18.2, 25.2, and 28.6~bars. Both measurement methods were performed alternately upon warming from the base temperature with a typical warming rate of 30~$\mu$K/hr. The temperature was measured by calibrated tuning fork (TF) thermometers \cite{Blaauwgeers2007JLTP1, Bradley2009JLTP1} below 0.6~mK and by a $^3$He melting curve thermometer above \cite{Zheng2016PRL1}. A magnetic field of 14~mT for a Pt NMR thermometer was applied to the superfluid in the direction perpendicular to the plane. For 28.6~bar, a cooldown of the superfluid in zero magnetic field was also performed. No significant difference was observed.

In the normal fluid or superfluid, with a low driving force, the damping force is proportional to the velocity of the MEMS plate. Therefore, the damping coefficient is independent of the velocity, and the spectra at various excitations can be normalized to a universal curve \cite{Gonzalez2013RSI1} \footnote{Also see Supplemental Material at [URL] for the linear spectra acquired in normal fluid $^3$He.}. However, when the driving force exceeds a threshold value in the superfluid, the velocity of the MEMS starts to deviate from the linear behavior. An excess damping emerges, and the MEMS-superfluid system enters a nonlinear regime where unusual behavior is observed.

Figure \ref{NLspectra} shows the resonance spectra, oscillator velocity \textit{versus} frequency, obtained at various excitations in the superfluid at 28.6~bar and 280~$\mu$K. One remarkable feature in the plot is the heavy distortion of the spectra around 5~mm/s. When normalized by the corresponding driving forces, the spectra do not overlap in the manner of linear spectra mentioned above. However, the low-velocity tails of the normalized spectra do collapse to a universal curve, indicating that the damping remains linear at low velocities. The excess damping mechanism does not set in until the plate velocity exceeds the threshold value. Beyond this point, the work done by the driving force does not increase the oscillator energy but is readily dissipated. Therefore, conventional mechanisms can not explain the observed nonlinear damping.
\begin{figure}
\includegraphics[width=0.75\linewidth]{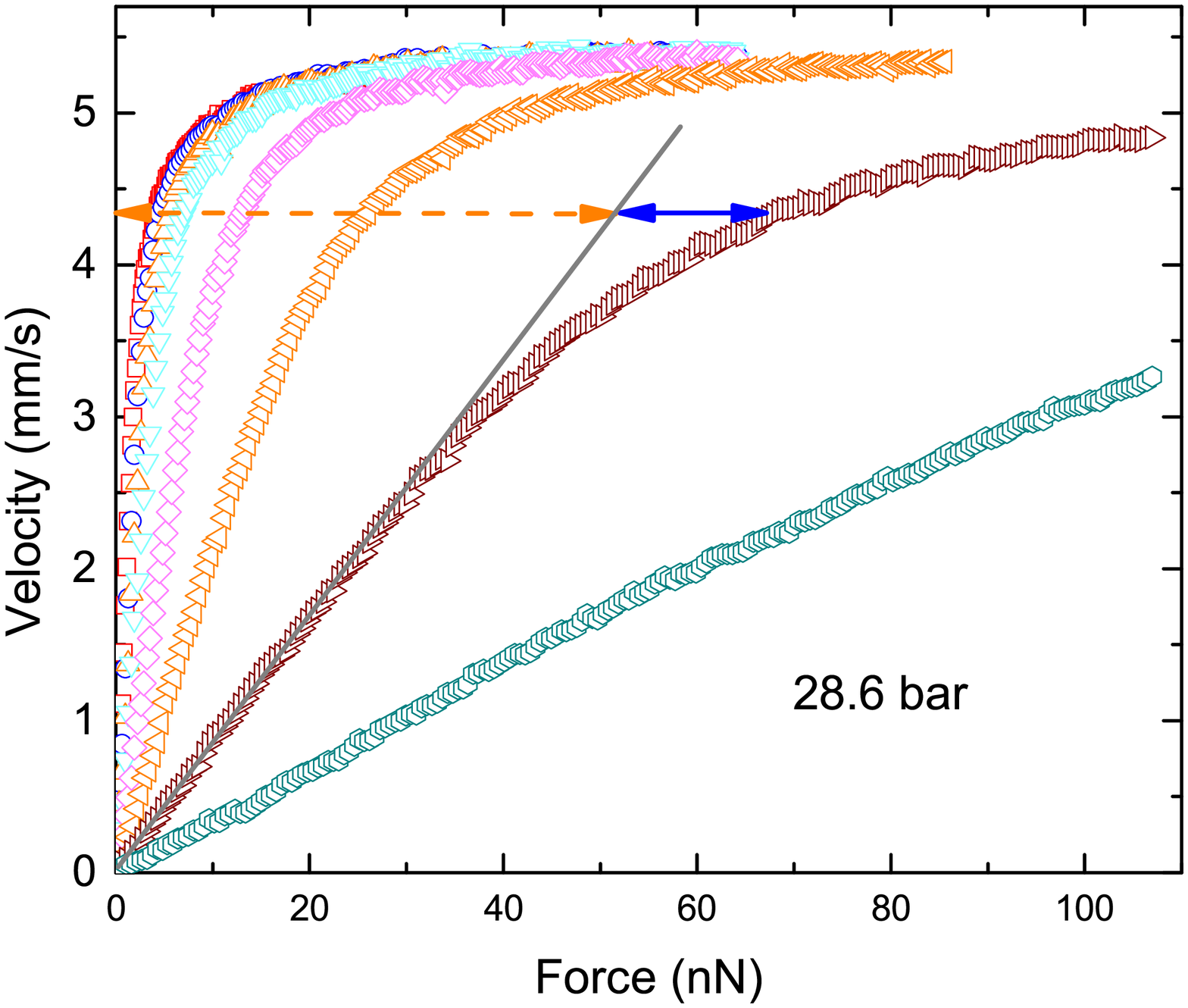}
\includegraphics[width=0.75\linewidth]{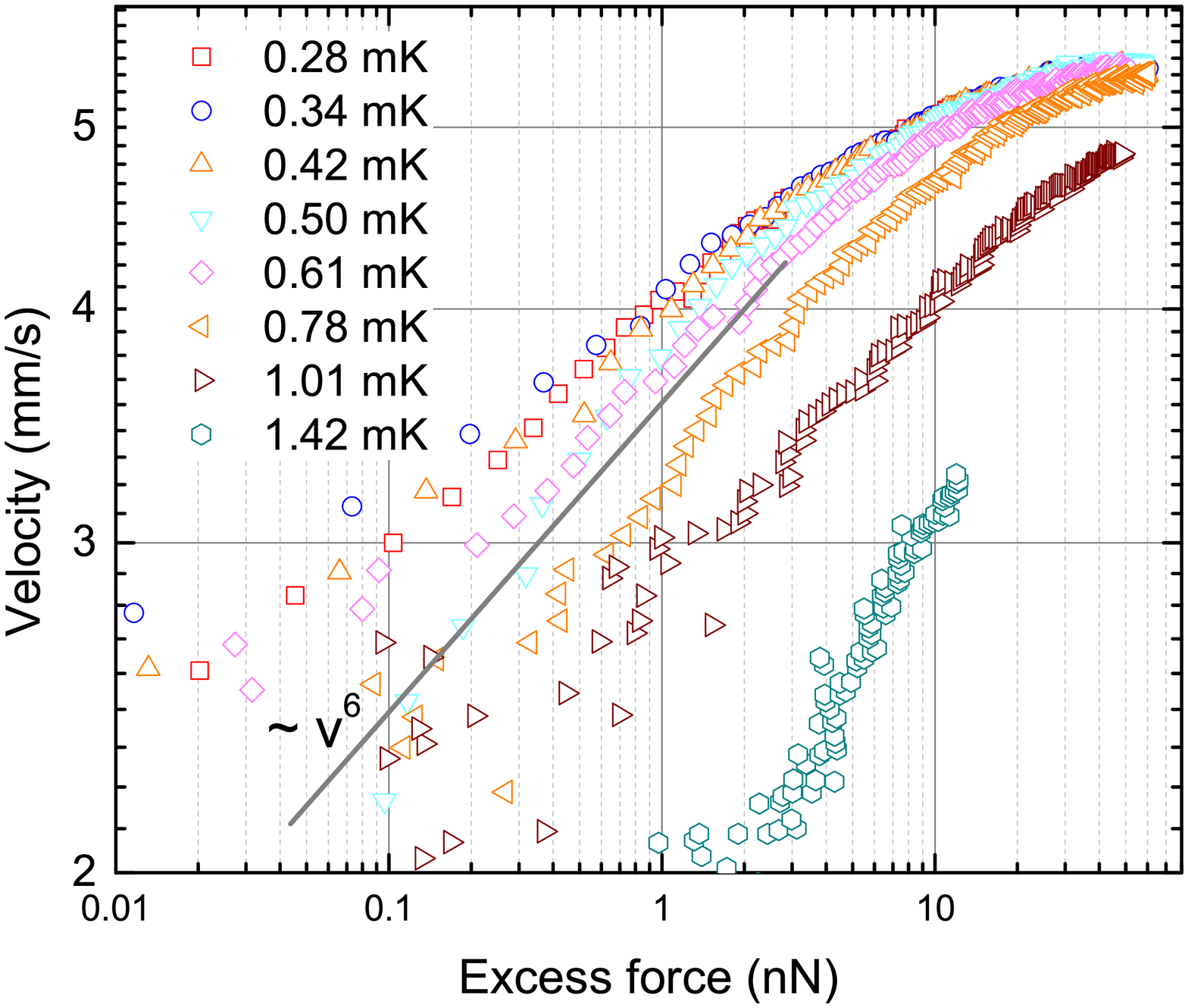}
\caption{The peak velocity of the oscillator against the driving force in a linear scale (\textit{top}), and against the excess damping force in a log-log scale (\textit{bottom}) at 28.6~bar for various temperatures. The dashed orange (solid blue) arrow represents the thermal (excess) damping at a given velocity. \label{NLVvsF}}
\end{figure}

The nonlinearity observed in superfluid $^3$He is characteristically different from what was observed in He\,II or in vacuum. In He\,II, the Lorentzian spectrum of a MEMS oscillator was deformed severely by the presence of a multi-peak structure \cite{Gonzalez2013JLTP1}. Strong hysteresis was also observed in He\,II, while the upward and downward sweeps in Fig.\,\ref{NLspectra} do not exhibit such hysteretic behavior \footnote{see Supplemental Material at [URL] for multiple upward and downward spectra at the same excitations.}. Similar nonlinear behavior was also observed in vibrating wires in He\,II and was interpreted as an interference from the bridged vortex lines connecting the surface of the oscillator and the boundary of the experimental chamber \cite{Hashimoto2007PRB1}. On the other hand, in vacuum at 4~K, the MEMS oscillator exhibits a typical Duffing type spectrum caused by the nonlinear electrostatic coupling of the comb drive \cite{Elshurafa2011JMS1} (see the inset of Fig.\,\ref{NLspectra}).

Figure \ref{NLspectra} demonstrates that no measurable resonance frequency shift is observed with the increase of the driving forces. Furthermore, for high driving forces the spectrum is practically flat near the resonance. Therefore, the excitation sweep can be readily acquired by stepping the driving force at a fixed driving frequency. Figure \ref{NLVvsF} shows the plot of the peak velocity against the driving force, the velocity-force curve, at 28.6~bar and various temperatures. One may divide the velocity-force curve into three regions. At low velocities, the curve is a straight line whose slope is inversely proportional to the thermal damping coefficient, $\gamma_{th}$. The thermal damping force, $F_{th}=\gamma_{th}v$, is caused by the scattering of the thermal quasiparticles. Therefore, the slope decreases with temperatures, as shown in Fig.\,\ref{NLVvsF} \cite{Zheng2016PRL1}. As the driving force continues to increase, the curve starts to deviate from the linear behavior above a threshold velocity. The deviation in this intermediate section indicates the onset of an excess damping beyond the thermal damping. The excess damping increases rapidly and eventually keeps the velocity from increasing further, asymptotically approaching a velocity defined as the critical velocity, $v_c$. The critical velocity does not depend on the temperature for $T\lesssim 0.4T_c$, even though the slope of the linear section varies in this temperature range.
\begin{figure}
\includegraphics[width=0.75\linewidth]{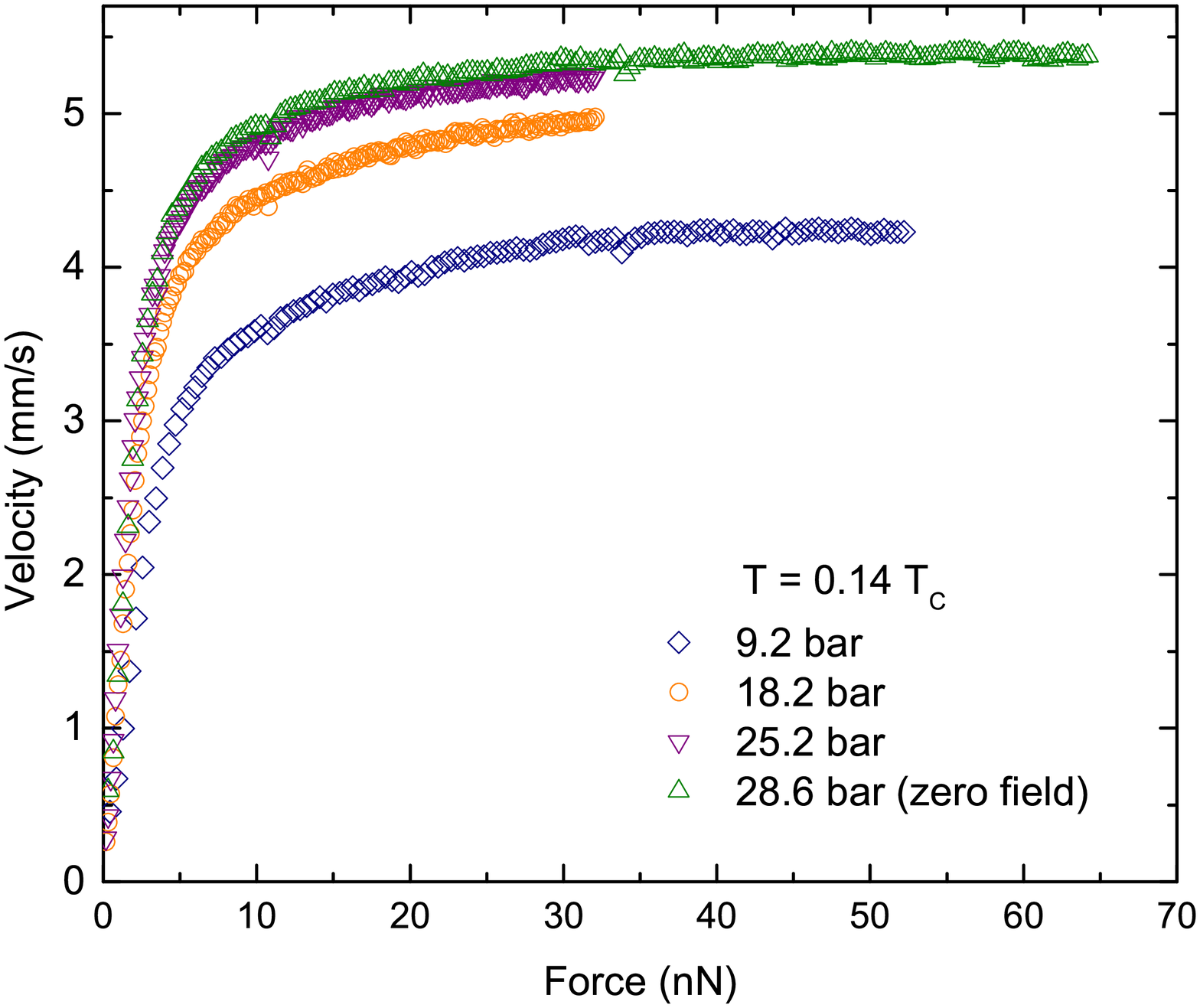}
\includegraphics[width=0.78\linewidth]{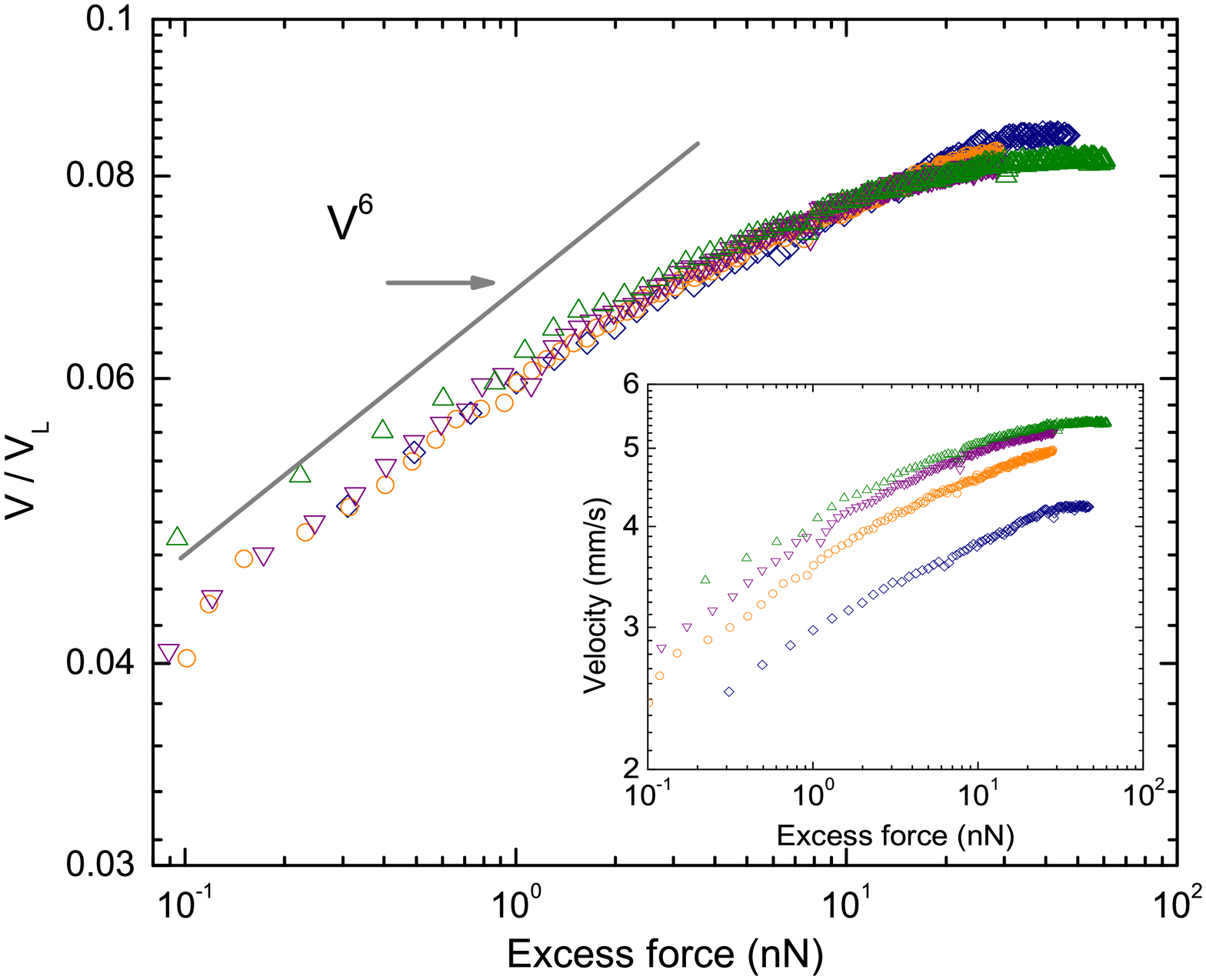}
\caption{The peak velocity of the oscillator against the driving force (\textit{top}) and the excess damping force (\textit{inset}) at $T = 0.14T_c$ for all pressures. The main \textit{bottom} panel shows the scaled velocity $v/v_{_L}$ against the excess damping force. The data at all pressures except for 28.6 bar (zero field) were taken at 14~mT. \label{NLVvsF4P}}
\end{figure}

Multiple studies have performed similar measurements using vibrating wires and quartz tuning forks, but observed a rather different behavior. Many of their oscillators experienced reduction of damping before the appearance of excess damping, in other words, bending upward rather than downward in the velocity-force plot shown in Fig\,\ref{NLVvsF} \cite{Castelijns1986PRL1, Carney1989PRL1, Bradley2009JLTP1}. The brief decrease in damping is followed by a rather fast change of the slope in the velocity-force curve in the direction of increasing damping. But the velocity-force curve for these devices does not show full saturation of velocity as observed in this work. The velocity where the abrupt slope change occurs was identified as the critical velocity. These studies found a consistent value of $v_c \approx v_{_L}/3$ for various vibrating wires and tuning forks. The initial decrease is now understood as a signature of the Andreev scattering of bulk quasiparticles in superfluid $^3$He \cite{Fisher1989PRL1}. The Andreev scattering correction becomes significant for $v\gtrsim k_{_B}T/p_{_F} \approx$ 4~mm/s at 250~$\mu$K, 28.6~bar. However, in Fig.\,\ref{NLVvsF}, there is no evidence of the Andreev scattering for the entire velocity range. We believe that unlike vibrating wires or tuning forks, the thin plate geometry of our oscillator would substantially weaken the condition for the Andreev scattering because the necessary potential barrier induced by the Doppler shift would not be effectively established near the plate undergoing shear motion. However, we cannot completely rule out the possibility that the excess damping mechanism kicks in prematurely to overshadow this effect.

The excess damping force, $F_{ex}$, can be separated from the total damping force, $F_t=F_{th}+F_{ex}$ (see Fig.\,\ref{NLVvsF}). The thermal damping (force), $F_{th}$, is inferred from the slope of the linear section. It is then subtracted from the total damping to yield the excess damping for each velocity. Figure \ref{NLVvsF} shows the velocity as a function of the excess damping in a log-log scale. In the low temperature limit, it is almost temperature independent and follows $F_{ex}\propto v^\sigma$ with $\sigma\approx 6$. A similar high order velocity dependence ($\sigma\approx 4$) was also observed in vibrating wires and quartz tuning forks \cite{Jackson2011thesis}, but could not be simply attributed to turbulence for which a $v^2$-dependence is expected \cite{Landaubookfluid}.

\begin{figure}
\includegraphics[width=0.75\linewidth]{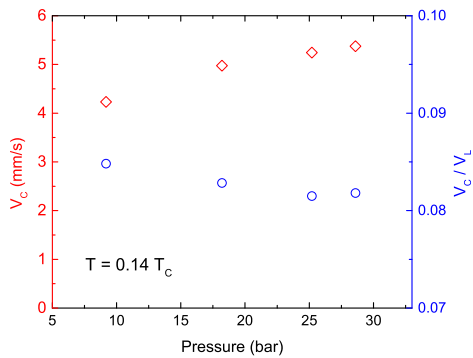}
\caption{The critical velocity (\textit{red diamonds}), $v_c$, and the ratio of the critical velocity to the Landau critical velocity (\textit{blue circles}), $v_c/v_{_L}$, against the pressure at $T = 0.14T_c$. \label{NLSVvsP}}
\end{figure}

The velocity-force curves around the base temperature for various pressures are plotted in Fig.\,\ref{NLVvsF4P}, displaying clear pressure dependences in the excess damping as well as the critical velocity. However, when the velocity is scaled by $v_{_L}$, the pressure dependence seems to disappear, suggesting that the pressure effect is likely inherited from the energy gap. It is fascinating to find the critical velocity in zero temperature limit is unusually low, $v_c\approx 0.08$\,$v_{_L}$ for all pressures (see Fig.\,\ref{NLSVvsP}).

The Landau critical velocity of a moving object should be determined by the maximum relative flow velocity, $v_{max}$, near the surface of the object. For an incompressible potential flow, $v_{max} = \alpha v_{ob}$ where $v_{ob}$ is the velocity of the moving object in the laboratory frame with a geometrical factor $\alpha$. For a cylindrical object moving in the perpendicular direction to its symmetry axis, a reasonable model for a vibrating wire, $v_{max} = 2v_{ob}$ at the top and bottom edge of the circular cross section. In contrast, for an ideal thin plate in the shear motion, $\alpha = 1$. A simple application of this fact would give $v_c = v_{_L}/\alpha$. However, Lambert made an intriguing proposal recognizing that the gapless surface states would be subjected to a flow of $v_{max}$ \cite{Lambert1990PB1, *Lambert1992PB1}. The quasiparticles can be generated near the surface by an infinitesimal amount of energy because of the closed gap and experiences the Doppler shift of $v_{max}p_{_F}$ rather than $v_{ob}p_{_F}$, which is the case for bulk fluid. He argued that a different type of dissipation -- not through pair-breaking process -- occurs at $v_c = v_{_L}/(1+\alpha)$ where the quasiparticles start to leak into the bulk because of the overlap of the spectra in energy. This is an attractive proposal since it naturally produces $v_c \approx v_{_L}/3$ for vibrating wires. Furthermore, Lambert proposed a quantum pumping mechanism, with which the critical velocity could be further reduced to a smaller fraction of $v_{_L}$ due to the fast reversal of the oscillating object \cite{Lambert1992PB1}. This mechanism requires the oscillation frequency $f > 35$~kHz at the saturated vapor pressure.

Our observation, $v_{c} < v_{L}/10$ at $f \approx 20$~kHz, cannot be fully explained by the mechanism described above. We do not believe that the massive loss of oscillator energy is related to vortices or other topological objects \cite{Winkelmann2006PRL1}, either. There was no noticeable hysteresis in the velocity-force measurement \footnote{see Supplemental Material at [URL] for the multiple upward and downward velocity-force sweeps at the same temperature.}; the multiple cooldowns produced practically identical results; the oscillator always recovered to the state before the intentional local heating, which would have disrupted the topological defects and objects.

We speculate that the unusually low critical velocity is directly related to the microscopic structure of the SABS. For a diffusive boundary, which is the case for this work, the surface states have an almost flat density of states (DOS) mid-gap band \cite{Nagato1998JLTP1, Murakawa2009PRL1}. This leads to a peculiar gap, referred to as the mini gap, in DOS between the upper edge of the band, $\Delta^*$, and the bulk continuum edge, $\Delta$. Quasiparticles excited into SABS are then promoted up to the edge of the mid-gap band by a multiple Andreev scattering process \cite{Zheng2016PRL1}. In the time scale of one oscillation ($\approx 50$~$\mu$s), it is estimated that $\sim 10^4$ scatterings off the oscillator wall would occur and effectively transfer the energy to the quasiparticles. This mechanism leads to anomalous low-temperature damping which was observed in our previous work \cite{Zheng2016PRL1}. In our oscillator geometry, the gap edges would experience the Doppler shifts, similar to the Lambert's process \cite{Lambert1992PB1}, $\Delta^* + v_{ob}p_{_F}$ and $\Delta - v_{ob}p_{_F}$ in the frame of reference of the oscillator. Therefore, the critical velocity for the quasiparticles in SABS to escape into the bulk would be $v_c = (\Delta - \Delta^*)/2p_{_F}$. According to quasiclassical calculations, $\Delta-\Delta^* \approx 0.2 \Delta$ \cite{Nagato2007JLTP1}, and consequently $v_c \approx 0.1 v_{_L}$, which is in a remarkable agreement with our result. This would also lead to the consistent critical velocities for $T<0.4T_c$ (see Fig.\,\ref{NLVvsF}), assuming that $\Delta-\Delta^*$ scales with $\Delta$. 

Theory predicts that the mini gap, $\Delta-\Delta^*$, shrinks quickly to zero and the low energy spectrum turns into a Dirac cone en route to the fully specular boundary \cite{Murakawa2009PRL1}. Therefore, we believe the critical velocity should be also sensitive to the boundary conditions, although it is not easy to envisage the trend without theoretical guidance. We do not believe that our observation is necessarily in contradiction to the recent result from the Lancaster group \cite{Bradley2016NP1}. It would be certainly interesting to investigate the uniform shear motion of a plate in various conditions of the surface.

In conclusion, using a MEMS oscillator in superfluid $^3$He we obtained an unusually low critical velocity $v_c \approx 0.08 v_{_L}$ for all pressures studied. We propose that this peculiar nonlinear behavior is directly related to the microscopic structure of the SABS near a diffusive boundary.

\begin{acknowledgments}
We would like to acknowledge Peter Hirschfeld, Anton Vorontsov, and Errki Thuneberg for helpful discussions, and the Lancaster Low Temperature group for providing custom-made quartz tuning forks used in this work. This work is supported by the National Science Foundation, Grant No. DMR-1205891.
\end{acknowledgments}

\bibliography{Reference}

\end{document}